\documentstyle[amssymb,epsfig,preprint,prb,aps]{revtex}
\tightenlines
\begin{document}
\title{Optimal packing of polydisperse hard-sphere fluids II}
\author{Ronald Blaak}
\address{University of Amsterdam, Faculty for Mathematics, Computer
Science, Physics and Astronomy, Kruislaan 403, 1098 SJ Amsterdam, The
Netherlands} 
\date{\today}
\maketitle

\begin{abstract}
We consider the consequences of keeping the total surface fixed for a
polydisperse system of $N$ hard spheres. In contrast with a similar
model (J. Zhang {\it et~al.}, J. Chem. Phys. {\bf 110},  5318
(1999)), the Percus-Yevick and Mansoori equations 
of state work very well and do not show a breakdown . For high
pressures Monte Carlo simulation we show three mechanically stable
polydisperse crystals with either a unimodal or bimodal particle-size
distributions. 
\end{abstract}

\section{Introduction}
\label{sec:intro}
Colloidal suspensions are never truly monodisperse, but are in general 
polydisperse. This might lead to unfavorable behavior, because it will
for example not be possible to form a high quality
crystal. Polydispersity will also have consequences on other
properties like the viscosity of the system. A better understanding of
these systems could therefore possibly lead to a handle to tune some
of the relevant properties of colloidal suspensions.

In a recent paper \cite{Zhang:1999JCP} we considered a system of $N$
hard spheres of which the total volume is fixed. The spheres are,
however, allowed to exchange volume under this constraint. Theory and
Monte Carlo simulations showed that the Percus-Yevick (PY) theory,
which normally is well suited for describing a polydisperse hard sphere
system, has a breakdown at a low volume fraction ($\eta \thickapprox
0.260$). According to the theoretical description, the
size-distribution function of the particles can not be normalized above
this threshold. In simulations it is found that above this packing
fraction a few of the particles will grow bigger on increasing
pressure. They will contain most of the available volume and will be 
 surrounded by a sea of small particles. The size of these big
particles becomes of the order of the simulation box, therefore finite
size effects do not allow to conclude how to describe the system at
high pressures. 

In this article we will focus on a similar system, but now the total
surface of the particles is fixed. In the Monte Carlo simulations we
performed, we allowed particles to exchange amounts of surface under
the constraint that the total surface remains constant. An
experimental system with this type of behavior, would be a system of
surfactant molecules, which are free to aggregate into spherical
micelles. For this model system we derive the Helmholtz free energy,
which is given by the free energy of a polydisperse system with an
arbitrary size-distribution, subject to two constraints, both the
number of particles and the total surface of the particles are
fixed. We define the optimal size-distribution of this system as the
one that minimizes this restricted Helmholtz free energy.  

The remainder of this paper is organized as follows: in Sec.
\ref{sec:theory} we derive the Helmholtz free energy of this system
and calculate the optimal size-distribution. In
Sec. \ref{sec:simulations} we show the results of our computer
simulations and make a comparison with the theoretical predictions.  
In Sec. \ref{sec:crystal} we discuss the possibilities to form a high
density crystalline phases and conclude in Sec. \ref{sec:conclusion}
with a discussion of our results.

\section{theory}
\label{sec:theory}
In a multicomponent system the ideal entropy is given by a simple
expression
\begin{equation}
-Nk_B\sum_iw_i\ln(\Lambda_i^3\rho w_i),
\label{eq:ideal}
\end{equation}
where $w_i$ is the molar fraction of species $i$ and $\Lambda_i$ its
thermal wavelength. In a true polydisperse system, however, this
entropy is infinite \cite{Salacuse:1982JCP}. Normally one would
describe such a system 
by distributing the particles over boxes according to some property,
e.g. diameter or volume, which would enable us to distinguish them.
In the case of a fixed particle-size distribution this is perfectly
allowed for whatever property one wishes to choose. In our case,
however, the particle-size distribution is allowed to change and as a
consequence the labeling property will influence the theoretical
description of the system \cite{Zhang:1999JCP}. The choice of how to
label the different boxes is in fact determined by an a priori
probability assumption, which in this case is dictated by the way
polydispersity is sampled.  

The way we sample the polydispersity can be thought of to describe a
system with a large, constant  number of tiny particles, which form
$N$ spherical aggregates. Exchanging surface between the aggregates
would then imply exchanging some of these tiny particles. The number
of these particles forming the aggregate is proportional to the
surface, which forms a natural labeling of our system.

The fact that the total surface of the $N$ spherical particles in
conserved leads to a natural length scale
$\langle\sigma^2\rangle^{(1/2)}$, where $\sigma$ is the diameter of a
particles and $\langle\cdot\rangle$ denotes an average over
particle-size distributions. We introduce the reduced pressure 
${P}^{*}=(k_{B}T)^{-1}P \langle\sigma^{2} \rangle^{(3/2)} $, where
$k_{B}T$, the Boltzmann constant times the temperature $T$, is our
unit of energy.

For an $n$-component hard sphere mixture, the compressibility pressure
in the  Percus-Yevick approximation yields \cite{Salacuse:1982JCP}:
\begin{equation}
\label{eq:PY-eos}
\frac{\pi }{6}P^*=\frac{\xi _{0}}{1-\xi _{3}}+\frac{3\xi _{1}\xi
_{2}}{(1-\xi_{3})^{2}}+\frac{3\xi _{2}^{3}}{(1-\xi _{3})^{3}},
\end{equation}
where the $j$-moment of the particle-size distribution $\xi _{j}$ is
defined as 
\begin{equation}
\label{eq:xi-j}
\xi _{j}=\frac{\pi}{6}\sum_{i}\rho _{i} \sigma_{i}^j,
\end{equation}
where the index $i$ is used to denote the different particle species,
$\rho _{i}=N_{i}/V$ is their number density, and $\sigma_{i}$ is their
diameter.

Equation (\ref{eq:PY-eos}) is also valid for a continuous size-distribution, in
which case the sum in Eq. (\ref{eq:xi-j}) is replaced by an integral. The
corresponding expression for the chemical potential of a species with
diameter $\sigma$ is\cite{Baxter:1970JCP} 
\begin{equation}
\label{eq:PY-mu} 
\mu^* =\ln\left[\rho \Lambda ^{3}W(s)\right]-
\ln (1-\xi _{3})+\frac{3\xi _{2}\sigma}{(1-\xi _{3})}+\frac{3\xi
_{1}\sigma^{2}}{(1-\xi _{3})}+\frac{9\xi_{2}^{2}\sigma^{2}}{2(1-\xi
_{3})^{2}}+\frac{\pi}{6}P^* \sigma^{3},
\end{equation}
where $\Lambda $ is the de-Broglie thermal wavelength, $\sqrt{h^{2}/(2\pi
mkT)}$, and $W(s)$ is the probability density to find a particle with a
surface around $s= \pi \sigma^2$. The pressure $P^*$ is given by
Eq. (\ref{eq:PY-eos}). 

In an (NPT) description, the Gibbs free energy ${\cal G}$ of the system
fulfilling the constraints, must be at a minimum, and is given by the
following functional
\begin{equation}
{\cal G}[W(s)] = \int \mu^* W(s) ds - {\cal L}_0 \int W(s) ds - {\cal
L}_1 \int s W(s) ds.
\end{equation}
The Lagrange multipliers ${\cal L}_0$ and ${\cal L}_1$ ensure that the
conservation of the number of particles and of the total amount of
surface respectively are satisfied. For a minimum of the Gibbs free
energy the functional derivative of ${\cal G}$ with respect to $W(s)$
should be zero, which implies that $W(s)$ must be of the form:
\begin{equation}
\label{eq:pdf}
W(s)=\exp \left\{ \sum_{i=0}^{3}\alpha _{i} \sigma^{i}\right\} ,  
\end{equation}
where the values of $\alpha_{1}$ and $\alpha_{3}$ are explicitly given by
\begin{eqnarray}
\alpha _{1} &=&-\frac{3\xi _{2}}{1-\xi _{3}},  \label{eq:alpha-1} \\
\alpha _{3} &=&-\frac{\pi}{6} P^*.  \label{eq:alpha-3}
\end{eqnarray}
The coefficients $\alpha _{0}$ and $\alpha_{2}$ are determined by the
constraints on the number of particles and the total of available
surface. Note that all $\xi _{i}$ $(i=1,2,3)$ are positive. Moreover, $\xi
_{3}$ is equal to the volume fraction $\eta $, and is therefore necessarily
less than one. Hence, $\alpha_{1}$ and $\alpha_{3}$ are always negative.
This also implies that the particle-size distribution can always be
normalized. This is different from our previous model
\cite{Zhang:1999JCP}, where the constraint on the total volume implied
that $\alpha_{3}$ could be zero and gave rise to a critical
endpoint on the equation of state, beyond which the PY-theory is not
capable of describing that system. In this case, however, it is
$\alpha_2$ which can change sign and become positive due to the
constraint on the surface. This will lead to the formation of bimodal
particle-size distributions in this model. An argument similar to the
one used in Ref. \cite{Zhang:1999JCP} to show the break down of the
Percus-Yevick equation of state, can also be used here to show that no
break down will occur in this case, where the total surface of the
particles is constrained \cite{Trizac:1999PC}. 

\begin{center}
\begin{figure}[h]
\epsfig{figure=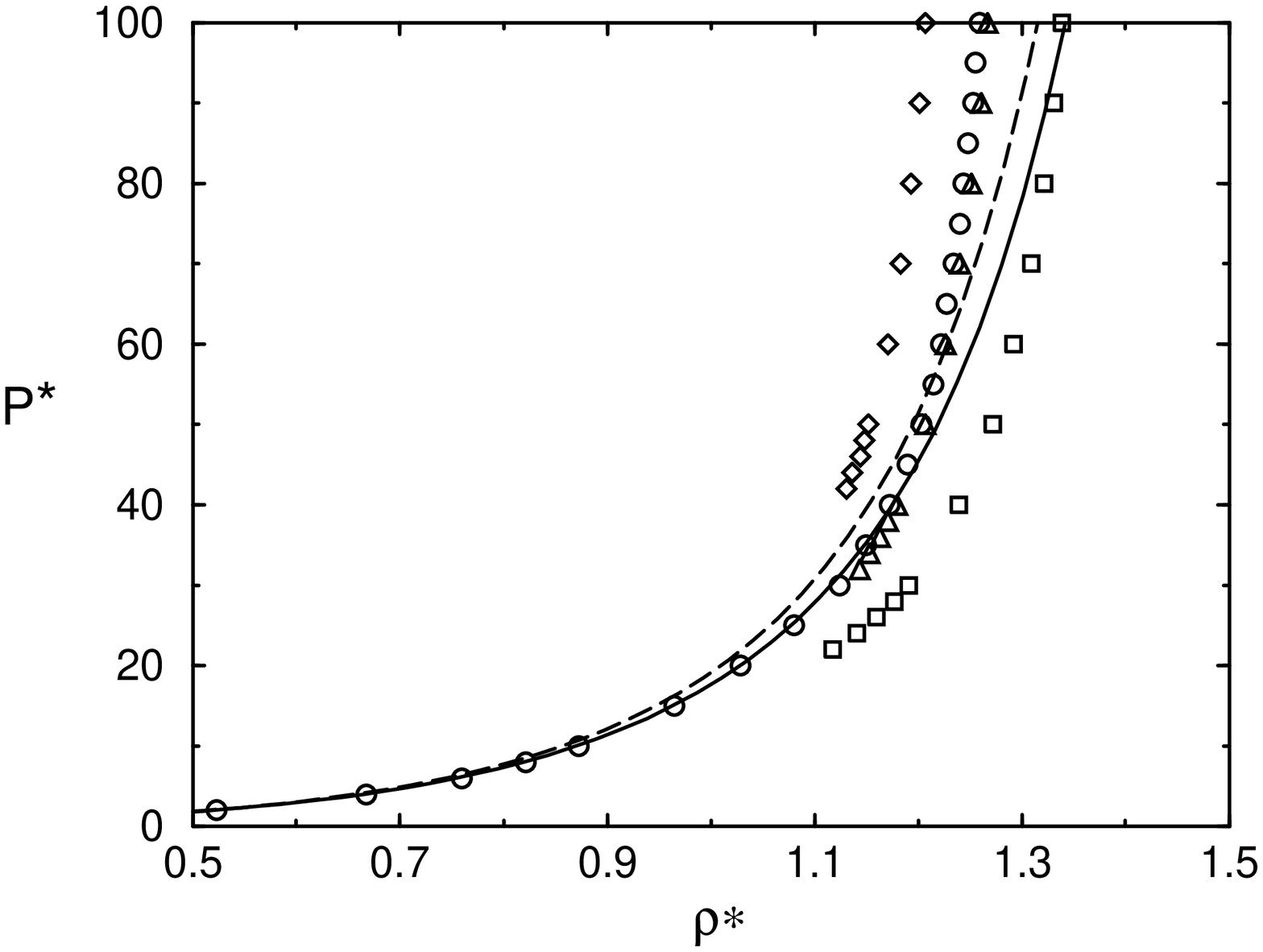,width=8.5cm,angle=0}
\vspace{3cm}
\caption[a]{\label{fig:eos-rho} Blaak, Journal of Chemical Physics}
\end{figure}
\end{center}

By self-consistently solving Eqns. (\ref{eq:alpha-1}),
(\ref{eq:alpha-3}) and the two equations imposed by the constraints,
we obtain the equation of state of this model (Figs. \ref{fig:eos} and
\ref{fig:eos-rho}) and the particle-size distributions in equilibrium
according to the PY-theory. The equation of state is presented here
both as function of the packing fraction $\eta$ and reduced density
${\rho}^{*}=(N/V) \langle\sigma^{2} \rangle^{(3/2)}$, because
in contrast with most simulations the volume of the particles involved
is not constant.

In the limit where the density goes to zero $\alpha_1=\alpha_3=0$ and
the particle-size distribution function $W(s)$ is an exponential
decaying function. For increasing densities the constraint on the
total surface will lead to an increasing value of $\alpha_2$, which
becomes zero at a packing fraction $\eta = 0.2169$ and reduced
pressure $P^* = 0.7025$. In order for the particle-size distribution
function to become bimodal, however, it should have a local
minimum. This requires its derivative with respect to the surface to
be zero 

\begin{equation} 
\frac{d W(s)}{d s} = W(s) \left( \frac{\alpha_1}{2 \sigma} + \alpha_2
+ \frac{3 \alpha_3 \sigma}{2} \right) = 0.
\end{equation} 
This point, which is characterized by the equation $\alpha_2^2 = 3
\alpha_1 \alpha_3$, can be numerically evaluated and 
leads to a critical value of the density $\eta_C = 0.3889$
and pressure $P^* = 3.4019$ above which bimodal behavior can be
observed. 

For the more accurate equation of state provided by Mansoori {\it et
al.} \cite{Mansoori:1971JCP}, the pressure should be replaced by
\begin{equation}
\label{eq:HS-eos}
\frac{\pi }{6}P^*_{HS}=\frac{\xi _{0}}{1-\xi _{3}}+\frac{3\xi _{1}\xi
_{2}}{(1-\xi_{3})^{2}}+\frac{3\xi _{2}^{3}}{(1-\xi _{3})^{3}} -
\frac{\xi_3 \xi _{2}^{3}}{(1-\xi _{3})^{3}}.
\end{equation}
This equation of state is obtained by $P^*_{HS} = \frac{2}{3} P^* +
\frac{1}{3} P^*_{Vir}$, where $P^*_{Vir}$ is the virial pressure. The
analysis is identical to that of the PY equation of state, but leads
to slightly different values. $\alpha_2$ becomes zero at $\eta =
0.2174$ and $P^*_{HS} = 0.7016$, and the critical point is found at
$\eta_C = 0.3944$ and $P^*_{HS,C} = 3.4362$.

The different equations of state, lead here not only to a different
pressure value for given density, but also to slightly different
particle-size distributions. This is in contrast with the fixed volume
model, where the different equations of state for given density would
only influence the value of the pressure.
  
\section{Simulations}
\label{sec:simulations}
For the Monte Carlo simulations we used the isothermal-isobaric 
ensemble or sometimes referred to as $NPT$-simulations
\cite{Book:Frenkel-Smit}. This means that the number of particles $N$, the
pressure $P$ and the temperature $T$ are fixed. In the simulations
there are three types of trial moves. The positions of the particles
are allowed to change by small amounts and we allow the simulation
box to fluctuate, in order to equilibrate with respect to the applied
pressure. In the case of the simulations of crystals we use a
rectangular box for which the three boxlengths are allowed to change
independently, while for the other simulations the shape of the 
simulation box remains cubic.

A third type of trial move is required in order to sample the
polydispersity. To this end we select at random two particles and
attempt to exchange an amount of surface between them, which is
uniformly drawn from  the interval $[-\Delta S_{max},\Delta
S_{max}]$. Here $\Delta S_{max}$ is the maximum amount of surface we
allow to be exchanged, and is adjusted such that the acceptance of this
move is between 35 and 50\%. Note that although the total amount of
surface of the particles is fixed the volume they will occupy will
change. 

The equation of state, resulting from our computer simulations of the
isotropic fluid performed on 1000 particles, is shown in
Figs. \ref{fig:eos} and  \ref{fig:eos-rho}. It compares nicely with
the equation of state of the Percus-Yevick approximation
(\ref{eq:PY-eos}). The equation of state of Mansoori {\em et al}
(\ref{eq:HS-eos}), performs even better. 

\begin{center}
\begin{figure}[h]
\epsfig{figure=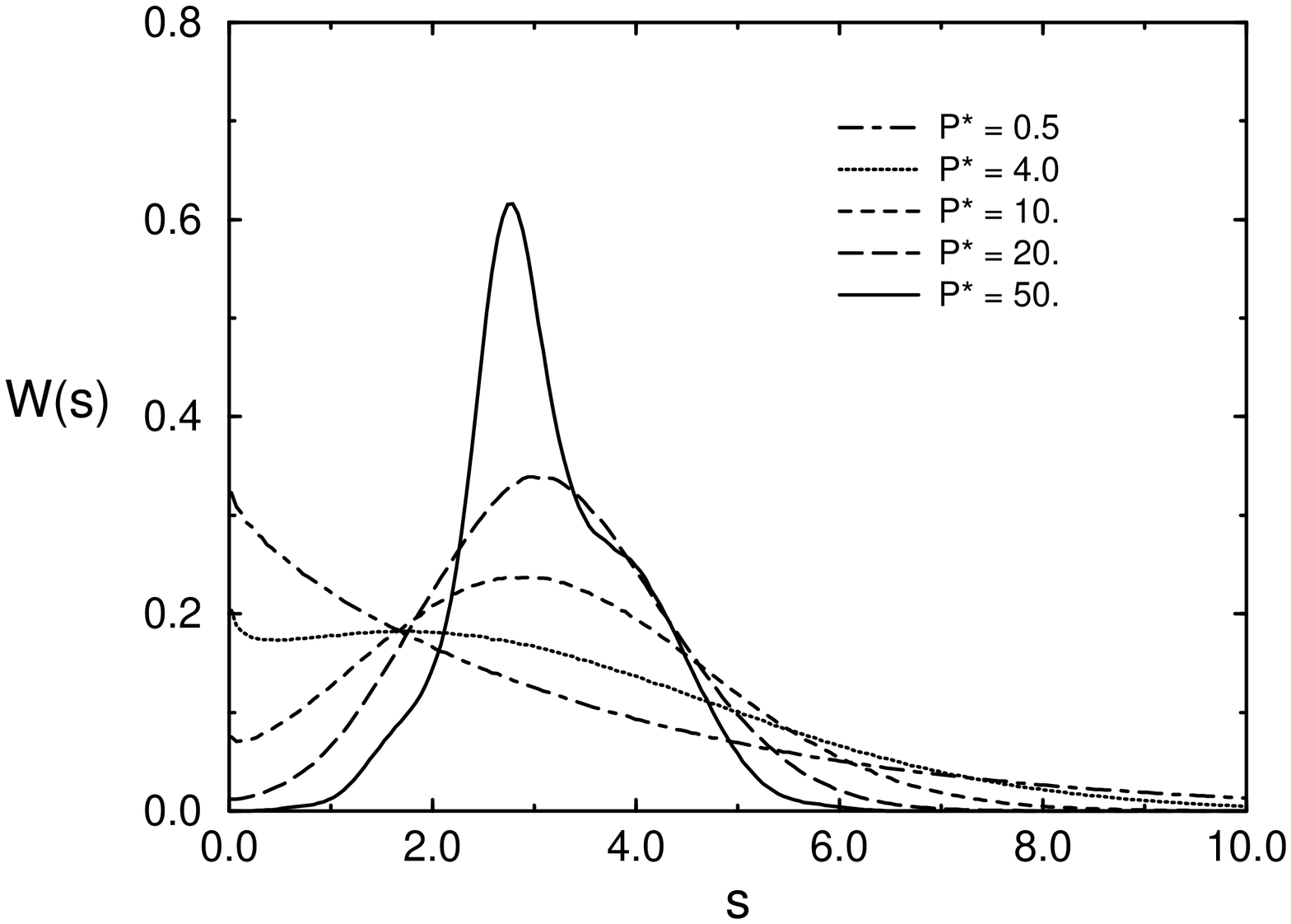,width=8.5cm,angle=0}
\vspace{3cm}
\caption[a]{\label{fig:pdf-iso} Blaak, Journal of Chemical Physics}
\end{figure}
\end{center}

\begin{center}
\begin{figure}[h]
\epsfig{figure=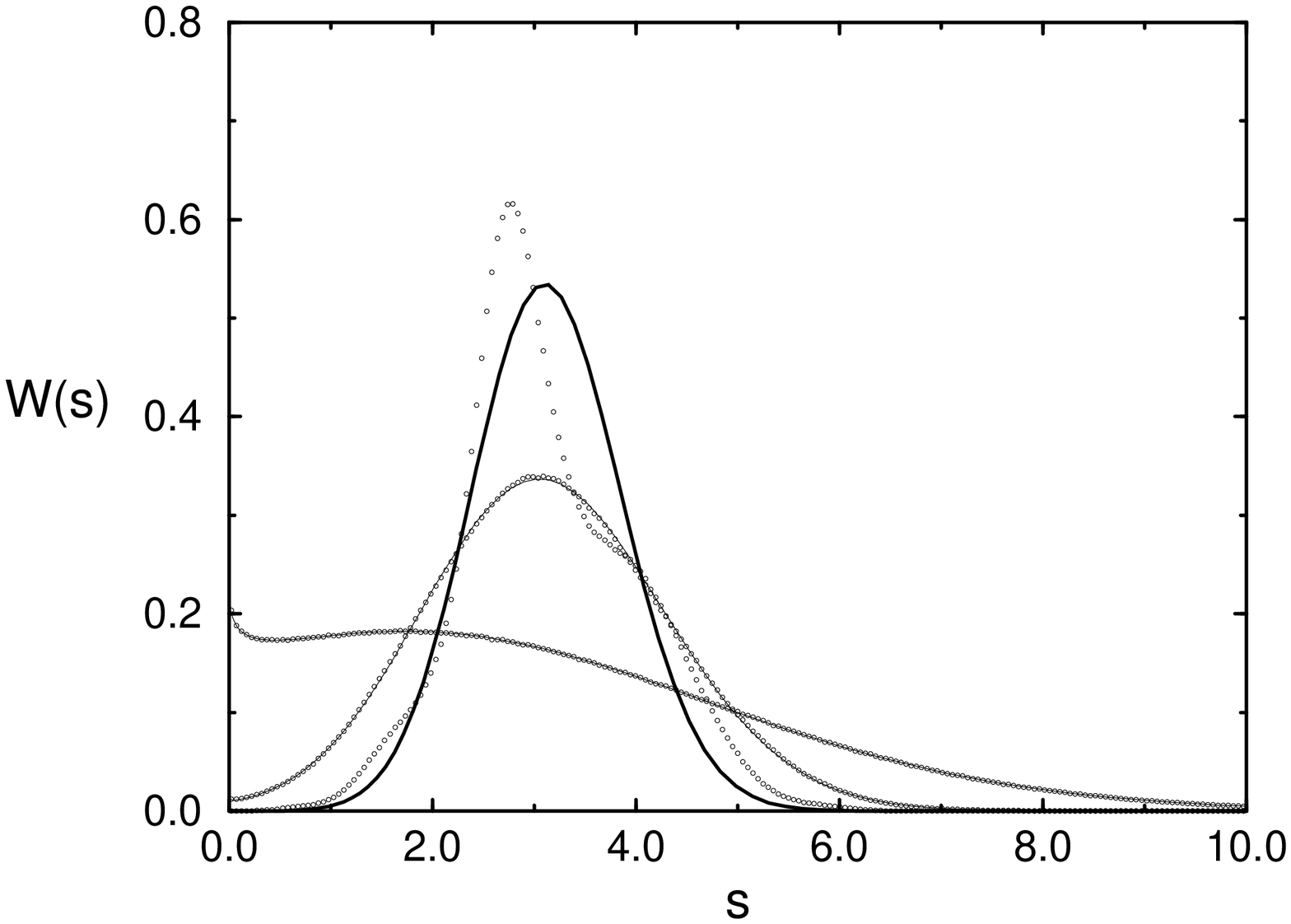,width=8.5cm,angle=0}
\vspace{3cm}
\caption[a]{\label{fig:pdf-com} Blaak, Journal of Chemical Physics}
\end{figure}
\end{center}

In Fig. \ref{fig:pdf-iso} we show several particle-size distributions
obtained from our simulations of the isotropic fluid. For 
low pressure we find an exponential decay. For the reduced pressure
$P^*=4.0$ we observe that it is bimodal, which is in agreement with
the theoretical predictions. For increasing pressure the fraction of
small particles decreases. Theoretically it remains bimodal, since
$\alpha_1$ in (\ref{eq:pdf}) is always negative, however it becomes
very small and in our simulations has disappeared at $P^* = 20.0$. In
Fig. \ref{fig:pdf-com}, a comparison is made between the measured and
predicted distributions. As can be seen they coincide nicely up to
$P^* = 20.0$ and $\eta=0.568$, and upon increasing pressure the
width of the unimodal distribution decreases.

This can be partially understood by the following argument. Upon
increasing pressure the system will try to adapt by forming a higher
density. This can be achieved by decreasing the volume of the
simulation box. One way to accomplish this would be by crystallizing. In
our simulation however there is another way in which this can be
achieved: decreasing the volume occupied by the particles. This is
possible, because the total surface of the particles is fixed, but
their total volume is free to change. If a special case of
the Power Mean inequality is applied to the particle diameters we obtain
\begin{equation}
^3\sqrt{\frac{1}{N} \sum \sigma_i^3} \geq \sqrt{\frac{1}{N} \sum
\sigma_i^2},
\end{equation}
where equality only holds if all diameters are the same. Therefore
the volume occupied by the particles can be lowered by forming a more
uniform size-distribution.

If the pressure is increased even more the size-distribution starts to
deviate again from the single peaked distribution
(Fig. \ref{fig:pdf-iso}), and starts to 
develop second peaks. This suggest that the limit to what extend the
system can equilibrate under higher pressures by becoming more
uniform is reached. Although it is not observed within the duration
of our simulations, one possibility is that it will try to stabilize by
forming a crystalline structure with possible different sized particles. 

\section{Crystal phases}
\label{sec:crystal}
As mentioned in the previous section this system can equilibrate in
two ways to higher pressures, by forming a more uniform
size-distribution or by forming a more ordered structure. In order to
explore this last possibility we tried several possibilities. 

For a monodisperse system the highest density which can be obtained,
would be by forming a face-centered-cubic crystal (FCC) or the
hexagonal close-packed crystal (HCP). The highest packing fraction
$\eta$ which can be obtained by those crystals is $\eta
=\pi/\sqrt{18}=0.74048$.

In our case however we can also explore the possibilities to form
crystals with two or more particle sizes. One of these possibilities is
the formation of a simple cubic crystal (SC). By alternating two
particle sizes, as found in a NaCl-crystal, it can be easily shown
that the highest packing fraction which this $AB$ type crystal can reach
is $\eta=0.79311$. In this case the diameters in terms of the reference
particle are given by 1.30656 and 0.541196. Note that for this
calculation we have the additional constraint that the average
particle surface is fixed.

We also considered the $AB_2$ structure as observed for an AlB$_2$
crystal, where hexagonal closed packed layers of larger particles are
separated by hexagonal rings of smaller particles. The highest packing
fraction, which can be reached by this $AB_2$ type crystal, is 
$\eta=0.78211$ in the case that the particles have diameters 1.38828
and 0.73235.

\begin{center}
\begin{figure}[h]
\epsfig{figure=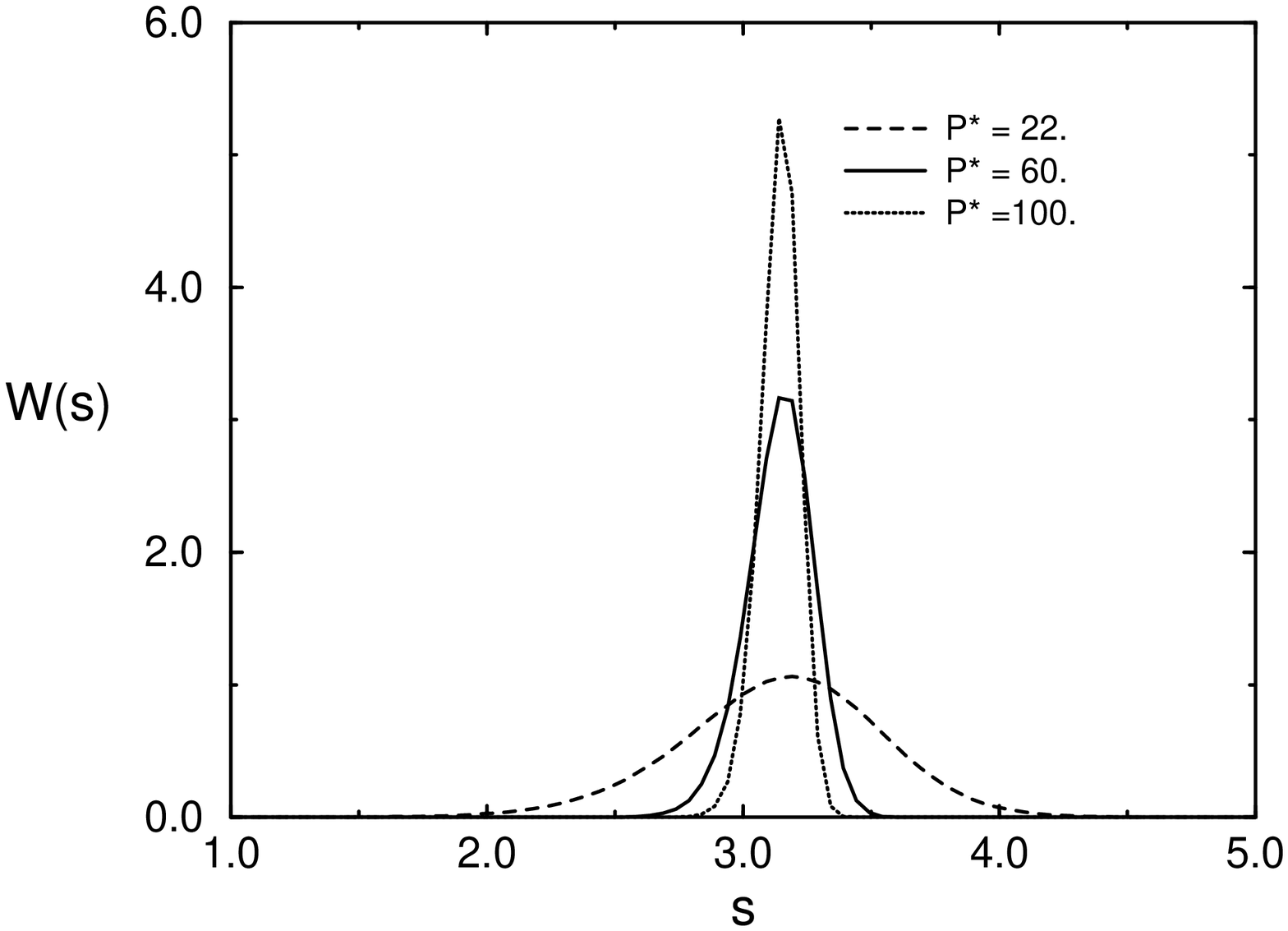,width=8.5cm,angle=0}
\vspace{3cm}
\caption[a]{\label{fig:pdf-fcc} Blaak, Journal of Chemical Physics}
\end{figure}
\end{center}

\begin{center}
\begin{figure}[h]
\epsfig{figure=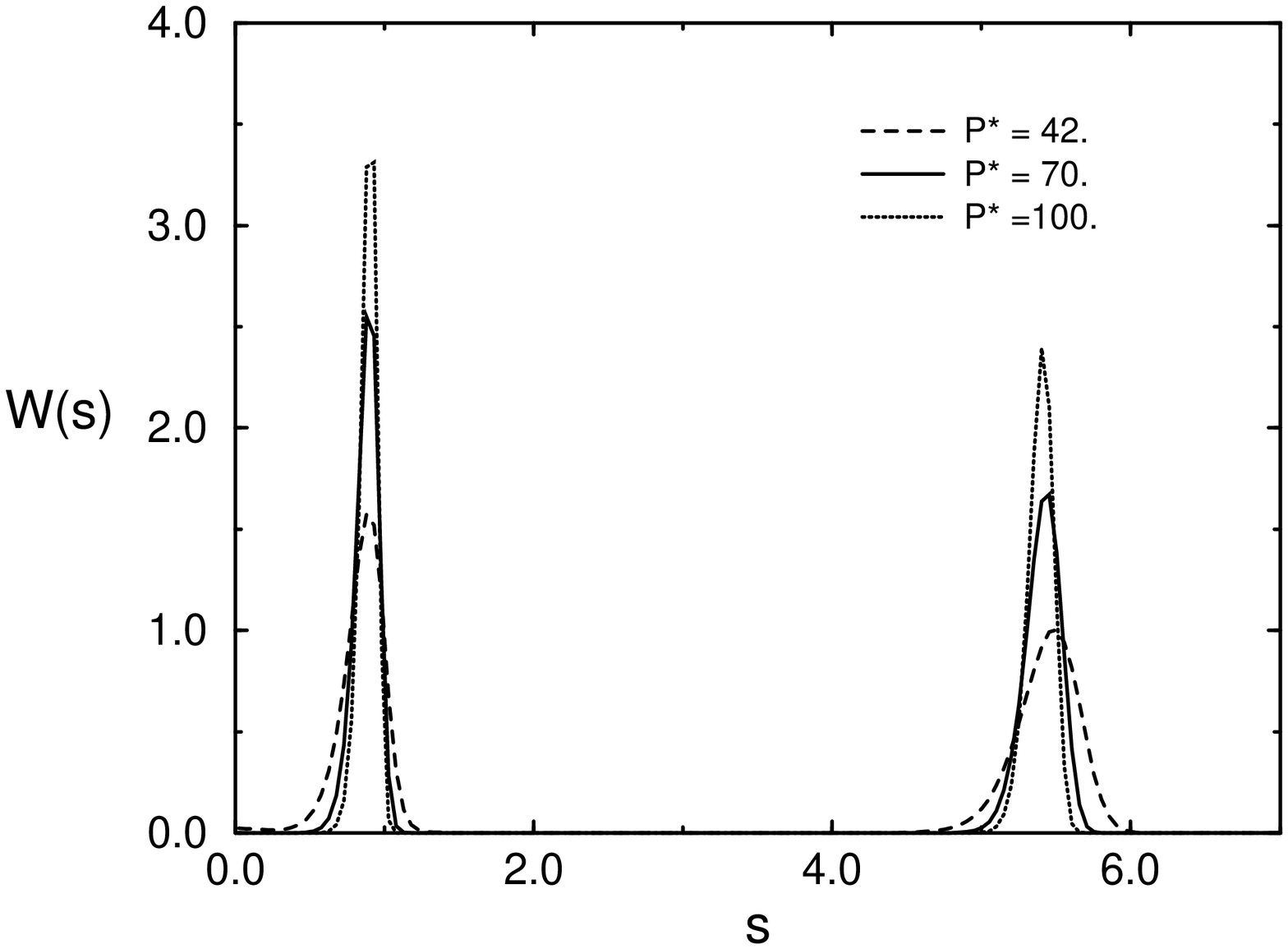,width=8.5cm,angle=0}
\vspace{3cm}
\caption[a]{\label{fig:pdf-ab} Blaak, Journal of Chemical Physics}
\end{figure}
\end{center}

\begin{center}
\begin{figure}[h]
\epsfig{figure=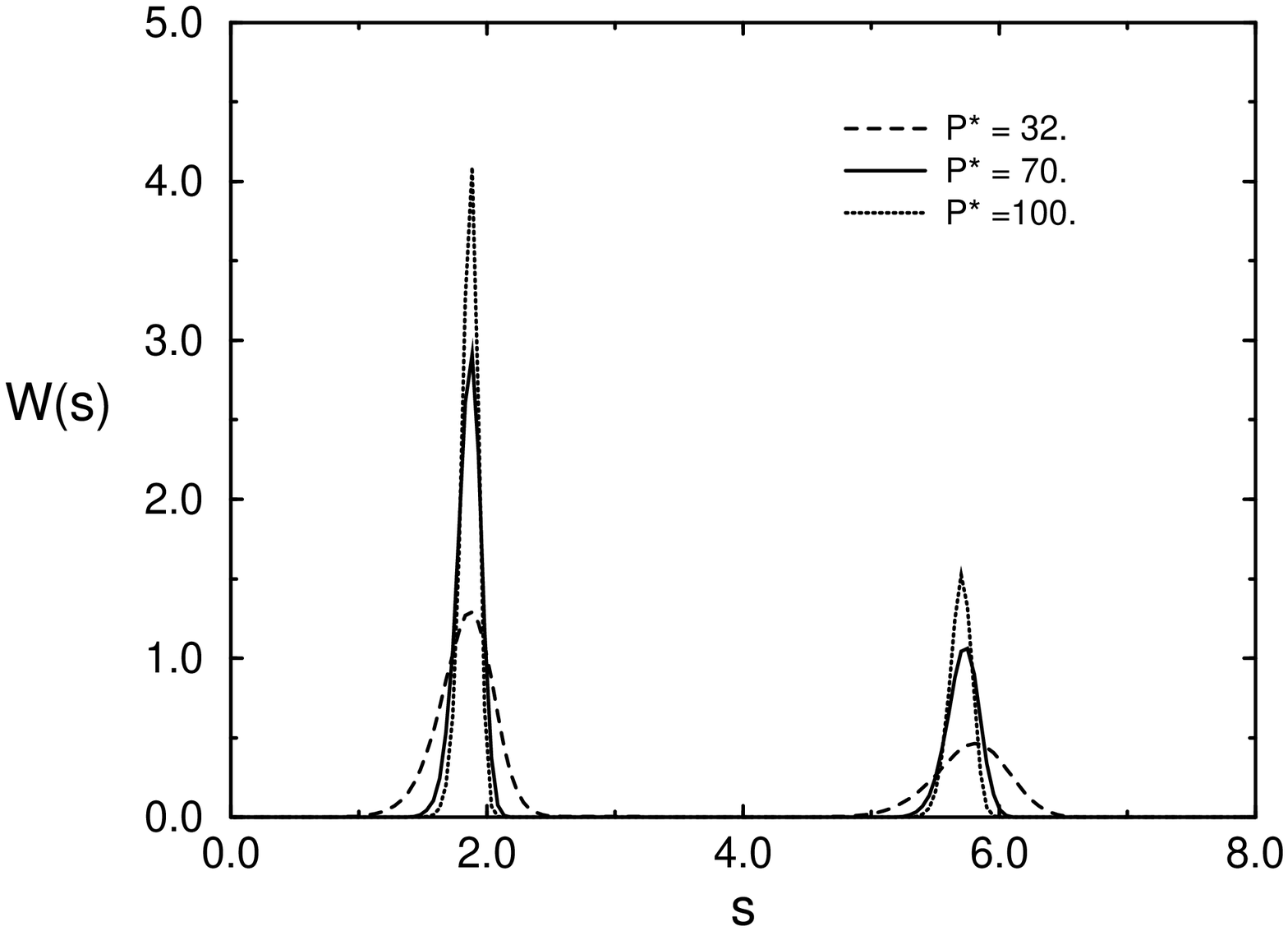,width=8.5cm,angle=0}
\vspace{3cm}
\caption[a]{\label{fig:pdf-ab2} Blaak, Journal of Chemical Physics}
\end{figure}
\end{center}

We performed simulations starting with these three perfect crystals,
using 720 particles for the FCC crystal, 1000 for the $AB$ crystal and
864 for the $AB_2$ crystal. For high pressures these crystals are all
at least mechanically stable. The equations of state of these branches
are also shown in Figs. \ref{fig:eos} and \ref{fig:eos-rho}. Note that
due to the non-fixed occupied volume of the particles a crystal with
higher packing fraction does not guarantee a higher number density.
The particle-size distribution (Figs. \ref{fig:pdf-fcc},
\ref{fig:pdf-ab},\ref{fig:pdf-ab2}) shows the broadening of the peaks 
if the pressure is lowered, and a shift of the maximum of the later two.
The lowest pressure points are close to the mechanical instabilities
of these crystals and give an lower bound to the coexistence pressure,
if any of these crystals is thermodynamically stable.

Compression on any of these crystal branches leaves the structure
intact, and the same equation of state is followed. However the
compression from the isotropic fluid does not show a transition to a
crystal branch. This might be due to the length of our simulation, and the
slow process of forming such a crystal from an unordered system. 

\begin{center}
\begin{figure}[h]
\epsfig{figure=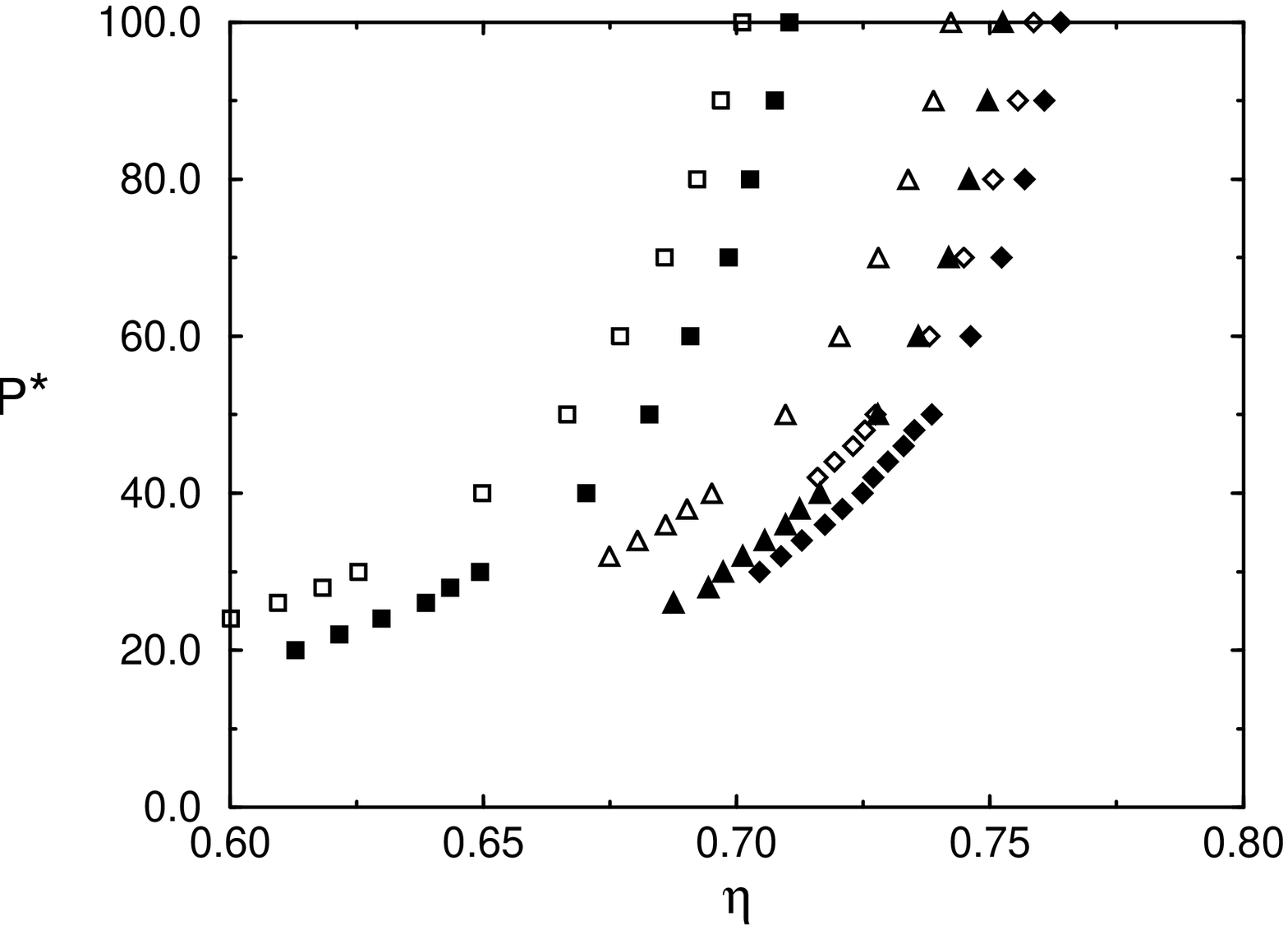,width=8.5cm,angle=0}
\vspace{3cm}
\caption[a]{\label{fig:eos-comp} Blaak, Journal of Chemical Physics}
\end{figure}
\end{center}

Although the particle-size distribution functions for the high density
crystals are almost monodisperse for the FCC crystal and bidisperse
for the $AB$ and $AB_2$ structures, it is not possible to use this in
order to simplify a cell theory to only one or two particle sizes
\cite{Cottin:1995JCP}. A comparison with simulation results of pure
mono- and bidisperse systems with the ``optimal'' sizes, reveals a
substantial difference in the equation of 
state (Fig. \ref{fig:eos-comp}). Therefore polydispersity should be
taken into account by incorporating the particle-size distribution
function in the cell theory, which falls outside the scope of this
article. As a consequence we can at this moment only conclude that
these crystals are mechanically stable, but not which of them is or
are thermodynamically stable.

We only considered the $AB$ and $AB_2$ compounds described above,
because they can lead to packing fractions beyond that of a
monodisperse FCC or HCP crystal. There are however more possible
candidates, like the CsCl and CaF$_2$ structures, or other compound
types like $AB_3$, $AB_4$, $AB_5$ and $AB_{13}$ \cite{Murray:1980PM}.
Even more complex crystals could be obtained by using three or more
particles sizes. A number of these structures are probably not stable,
because either the the packing fraction can not be high enough, or the
smaller particles are able to escape from their lattice positions
\cite{Cottin:1995JCP}.

\section{Conclusion}
\label{sec:conclusion}
We have investigated a polydisperse hard sphere system in which
particles can exchange surface under the constraint that the total
surface of the particles is fixed. In contrast with a similar model
\cite{Zhang:1999JCP}, the equation of state does not break down, and
predictions made by the approximation of Mansoori {\em et al} works
very well. 

For low pressures the particle-size distribution is an exponential
decaying function. On increasing pressure it becomes bimodal and
eventually it forms a single peaked distribution by eliminating the
smallest particles. If the pressure is increased even more there are
signs that it might become bimodal again, e.g. by forming a crystal
with two particle sizes. However, such a transition is not observed. 

We performed simulations on three types of polydisperse crystals: a
FCC-like crystal, an $AB$-crystal as found for NaCl, and an
$AB_2$-crystal as found for AlB$_2$. These crystals are all
mechanically stable, and are just the most basic types of crystals
possible. The polydispersity, however, influences the equation of
state such, that a simple cell theory can not predict the true
coexistence or thermodynamically stable crystal for these approximately
mono- and bidisperse crystals.

The formation of an ``Appolonian'' packing, which in principle can
lead to a space-filling structure, is not likely to happen, as the
constraint on the total surface prevents this. The study of this and a
previous model \cite{Zhang:1999JCP} have shown that these simple
models lead to surprising results. Although they are both somewhat
artificial we hope that they will stimulate the research towards more
realistic, related models. 

\section*{Acknowledgments}
We thank Daan Frenkel, Jos\'e Cuesta and Emmanuel Trizac for a
critical reading of the manuscript, as well as for their comments on
and discussions about the results presented here.

\newpage
\begin{center}
{\large FIGURE CAPTIONS}
\end{center}

\begin{center}
\begin{figure}[h]
\epsfig{figure=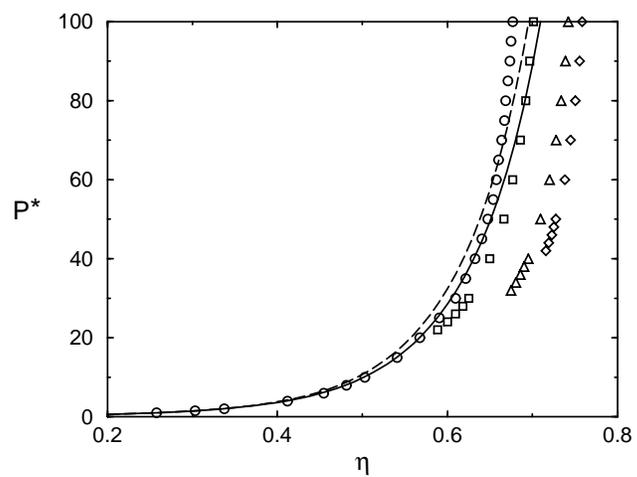,width=8.5cm,angle=0}
\vspace{3cm}
\caption[a]{\label{fig:eos} Blaak, Journal of Chemical Physics}
\end{figure}
\end{center}

\begin{enumerate}
\item
The equation of state of the polydisperse system. The circles are the  
simulation results of the isotropic fluid, the squares of an FCC
crystal, the diamonds of an $AB$ crystal, and the triangles of an $AB_2$
crystal. The theoretical predictions, Percus-Yevick (dashed) and the
Mansoori {\em et al} (solid), are also shown. 

\item
The equation of state of the polydisperse system as in
Fig. \ref{fig:eos}, but now the reduced pressure $P^*$ is given as
function of the reduced density $\rho^*$. Note that the differences are
caused by the fact that the occupied volume of the particles is not
fixed. 

\item
The particle distribution function for several reduced pressures
obtained by simulations of the isotropic fluid. At reduced pressure
$P^*=4.0$ the distribution is just bimodal. The highest pressure shown
($P^*=50.$) suggest the formation of a second peak.

\item
A comparison between the particle distribution function obtained from
simulation (points) and the Mansoori {\em et al} approximation
(lines), for reduced pressures $P^*=4.0$, $P^*=20.$, and
$P^*=50.$. There is a very good agreement, except for the highest
pressure where the predicted curve remains nicely single-peaked, while
the simulation suggest the formation of bimodal distribution.

\item
The particle distribution function for several reduced pressures on
the FCC branch. $P^*= 22.$ is the lowest pressure for which the
crystal did not melt during the duration of our simulation. 

\item
The particle distribution function for several reduced pressures on
the $AB$-crystal branch. $P^*= 42.$ is the lowest pressure for
which the crystal did not melt during the duration of our simulation.

\item
The particle distribution function for several reduced pressures on
the $AB_2$-crystal branch. $P^*= 32.$ is the lowest pressure for
which the crystal did not melt during the duration of our simulation.

\item
A comparison of the equations of state of polydisperse and mono- or
bidisperse systems. The squares correspond to a FCC crystal, the
diamonds to an $AB$ crystal, and the triangles to an $AB_2$ crystal. The 
open symbols correspond to the polydisperse systems, while the filled
symbols denote the results of true mono- or bidisperse
systems with fixed ``optimal'' size ratios. 

\end{enumerate}

\newpage

\newpage

\newpage

\newpage

\newpage

\newpage

\newpage

\end{document}